\documentclass[aps,pre,twocolumn,10pt]{revtex4-2}
\usepackage[dvipsnames]{xcolor}
\usepackage{graphicx} 
\usepackage{amsmath, amsfonts, amssymb,bm}

\usepackage{XCharter}
\usepackage[T1]{fontenc}
\usepackage{mathptmx}
\usepackage[framemethod=tikz]{mdframed}
\usepackage{amsmath,amsthm,amssymb,mathrsfs}
\usepackage{slashed}
\usepackage{graphicx}
\usepackage[export]{adjustbox}
\usepackage{float}
\usepackage{multirow, makecell}
\usepackage{hyperref}
\usepackage{enumitem}
\hypersetup{colorlinks=true,citecolor=red,linkcolor=NavyBlue,urlcolor=NavyBlue}
\usepackage[caption=false]{subfig}
 
\usepackage{natbib}

\usepackage{relsize}
\usepackage{footmisc}
\usepackage[left=1.55cm,right=1.55cm,top=1.6cm,bottom=1.56cm]{geometry}
\usepackage{tcolorbox}
\usepackage{longtable}
\usepackage{xcolor}

\usepackage{url}

\newcommand{\Asutosh}[1]{\textcolor{blue}{#1}}

\begin{document}

%\preprint{APS/123-QED}

\title{Synchronization in Networks of Heterogeneous Kuramoto–Sakaguchi Oscillators with Higher-order Interactions}% Force line breaks with \\

\author{Asutosh Anand Singh ${}^{1}$, Hiroshi Kori ${}^{2}$, and Chandrakala Meena ${}^{1}$}
\affiliation {${}^{1}$ Physics Department, Indian Institute of Science Education and Research (IISER), Pune, 411008, Maharashtra, India.}

\affiliation{${}^{2}$ Department of Complexity Sciences and Engineerings, Graduate School of Frontier Sciences, The University of Tokyo, Japan.}

%  \altaffiliation[Also at ]{Department of Physics, IISER Pune, Maharashtra, India}%Lines break automatically or can be forced with \\
% \author{Chandrakala Meena}%
%  \email{chandrakala@iiserpune.ac.in}
% \affiliation{%
%  Department of Physics, IISER Pune, Maharashtra, India
% }%

\date{\today}% It is always \today, today,

\begin{abstract}
How do the combined effects of phase frustration, noise, and higher-order interactions govern synchronization in globally coupled heterogeneous Kuramoto oscillators? To address this question, we investigate a globally coupled network of Kuramoto–Sakaguchi oscillators that includes both pairwise ($1$-simplex) and higher-order ($2$-simplex) interactions, together with additive stochastic forcing. Systematic numerical simulations across a broad range of coupling strengths, phase-lag values, and noise intensities reveal that synchronization emerges through a nontrivial interplay among these parameters. In general, weak frustration combined with mutually reinforcing coupling promotes synchronization, whereas strong frustration favors coherence under repulsive coupling. Increasing noise progressively suppresses coherence, smoothening the transition from desynchronized to synchronized states.
Forward and backward parameter sweeps reveal the coexistence of synchronized and desynchronized states. The presence and width of this bistable region depend sensitively on phase frustration, noise intensity, and higher-order coupling strength, with higher-order interactions significantly widening the bistable interval. To explain these behaviors, we employ the Ott–Antonsen reduction to derive a low-dimensional amplitude equation that predicts the forward critical point in the thermodynamic limit, the backward saddle-node point, and the width of the bistable region. Higher-order interactions widen this region by shifting the saddle-node point without affecting the forward critical point. Further analysis of Kramer's escape rate explains how noise destabilizes co-existence states and diminishes bistability. Overall, our results provide a unified theoretical and numerical framework for frustrated, noisy, higher-order oscillator networks, revealing that synchronization is strongly influenced by the combined action of phase frustration, stochasticity, and both pairwise and higher-order interactions.
\end{abstract}

\maketitle

%\tableofcontents

\section{\label{sec:level1} Introduction}

The emergence of synchronization in large networks of coupled oscillators stands as a central theme in nonlinear dynamics and complex systems, with wide-ranging applications in physics, neuroscience, engineering, and the social sciences \cite{strogatz2000kuramoto, buck1988synchronous,rungta2017network, acharyya2024master}. The term "synchronization" derives from the Greek words "syn" (meaning "together") and "chronos" (meaning "time") \cite{GHOSH2023114237}, reflecting the alignment of oscillatory rhythms, either among interacting components or with an external driving force.

Synchronization has been recognized as a ubiquitous phenomenon across both natural and engineered systems, encompassing areas such as biological populations \cite{buck1988synchronous} and systems {\cite{mehrabbeik2023impact, wang2024multi}}, climate \cite{meena2017effect}, power grids, and communication networks \cite{PhysRevE.100.062306}. Depending upon system properties and couplings, synchronization can manifest in various forms, including phase\cite{HO200243, ANISHCHENKO2002469}, frequency\cite{berner2019multiclusters}, and amplitude synchronization \cite{ZHANG2004442,meena2016chimera}. A major milestone in the study of collective synchronization was achieved with the introduction of the Kuramoto model, which describes a population of globally coupled phase oscillators interacting through sinusoidal functions of their phase differences \cite{winfree1980, Kurmaoto1984}. Despite its apparent simplicity, the Kuramoto model captures a nonequilibrium phase transition from incoherence to collective synchrony when the coupling strength exceeds a critical threshold \cite{STROGATZ20001, RevModPhys.77.137}. Since then, this framework has been widely applied across diverse domains, including Josephson junction arrays \cite{SWIFT1992239, PhysRevLett.76.404}, ensembles of neurons \cite{ramasamy2022effect, FARRERAMEGCHUN2025116368}, and firefly flashing \cite{buck1988synchronous}, cricket chirping \cite{walker1969acoustic}. In the social sciences, Kuramoto-type models have been used to study crowd synchronization, such as people walking on bridges \cite{strogatz2005, PhysRevE.75.021110}. 

The low-dimensional reduction method introduced by Ott and Antonsen \cite{ott2008low} further enabled an elegant analytical study of such high-dimensional oscillator ensembles, making the Kuramoto framework a foundational model for understanding emergent synchronization in complex systems.

The Kuramoto model has been extended to incorporate more realistic features, such as nontrivial frequency distributions \cite{crawford1994amplitude, PhysRevE.79.026204}, heterogeneous network topologies \cite{moreno2004synchronization}, and higher harmonics \cite{daido1996multibranch}. Recent studies have highlighted that pairwise network dynamical frameworks are often insufficient, as many real systems involving simultaneous multi-body interactions are modeled using simplicial complexes and hypergraphs \cite{battiston2020networks, boccaletti2023structure}. This approach also leads to the definition of a useful global synchronization measure known as the Kuramoto–Daido order parameter \cite{daido1996multibranch}.  Incorporating such higher-order interactions has revealed abrupt synchronization transitions \cite{skardal2020higher}, and multistability \cite{dai2025higher}. 

A significant factor influencing the collective dynamics of coupled Kuramoto oscillators is frustration \cite{PhysRevE.93.062315}, typically introduced through a phase lag in the coupling function as described by the sine term in the Sakaguchi-Kuramoto (SK) oscillators ensemble. Coupled SK oscillators exhibit behaviors such as the coexistence of synchronized and desynchronized states known as chimera, in which coherent and incoherent oscillator groups emerge simultaneously, and partially synchronized states that interpolate between full order and complete incoherence ~\cite{crawford1994amplitude,kuramoto2002}.

Stochasticity is also unavoidable in engineered systems, such as \Asutosh{{\cite{PhysRevLett.109.064101, doi:10.1137/110851584, Cestnik2022ExactFR, Tnjes2020LowdimensionalDF}}}, arising from intrinsic variability or external disturbances \cite{Tanaka_2020,kostin2023synchronizationtransitionssensitivityasymmetry}. Noise can destabilize coherence by inducing random phase slips or, under certain circumstances, favor synchronization through noise-induced ordering. The net effect depends sensitively on the noise intensity and the underlying coupling structure of the system~\cite{goldobin2005synchronization}.

Over the past several decades, numerous studies have investigated the separate roles of phase frustration, stochasticity \cite{PhysRevE.108.064124, PhysRevLett.120.264101}, and higher-order interactions in shaping synchronization. Yet their joint impact, especially in systems with explicit multi-body coupling terms, remains poorly understood. A unified framework that simultaneously accounts for phase frustration, noise, and higher-order coupling is still missing but has been explored in the case of pairwise coupling \cite{vukadinovic2025traveling, 10.1063/1.5053576}.

In this work, we address this gap by investigating a globally coupled network of Kuramoto–Sakaguchi oscillators augmented with both pairwise ($1$-simplex) and higher-order ($2$-simplex) interactions under stochastic forcing. Through numerical simulations and the Ott–Antonsen (OA) reduction, we demonstrate how the interplay between phase frustration, noise, and higher-order coupling profoundly affects synchronization. Our results demonstrate that higher-order interactions can accelerate the onset of synchronization, expand the bistable region, and mitigate the desynchronization effects of frustration and noise. We analyze the resulting multistability using the reduced amplitude equation and further quantify the stability of metastable states using Kramer's escape theory, which explains how noise modulates hysteresis width. Overall, this study presents a unified theoretical and computational framework for synchronization in oscillator networks, where the combined effects of frustration, stochasticity, and higher-order interactions are crucial for modeling real-world complex systems.

Our paper is organized as follows: Section~\ref{sec:level2} introduces the dynamical framework and presents the numerical results obtained from simulations. Section~\ref{sec:level3} provides the analytical results derived via the Ott–Antonsen reduction, followed by fixed-point analysis, stability characterization, bifurcation diagrams, and multistable states analysis using Kramers' escape rate. Finally, Section~\ref{sec:level4} summarizes and discusses the main findings of the study.

\section{\label{sec:level2}Model description and numerical simulations}

\begin{figure}
 \centering
    \includegraphics[width=0.8\columnwidth]{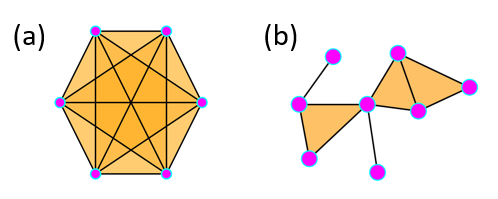}
    \caption{A representative architecture of the globally coupled oscillator system where the nodes ($0-simplexes$) are represented by magenta color, the pairwise connections, that is, the edges($1-simplexes$), are represented by black lines, and higher-order interactions ($2-simplexes$, filled triangles) are represented by orange color. Panel (a) represents all-to-all coupling, whereas panel (b) represents a smaller version of simplices.}\label{fig: Fig. 1}
\end{figure}

In our study, we consider the Sakaguchi-Kuramoto (SK) model \cite{PhysRevE.101.062213}, which is an extension of the Kuramoto model \cite{RevModPhys.77.137} through the inclusion of a phase lag term (\emph{frustration}) in the coupling function. We focus on a globally (all-to-all) coupled phase-oscillator system (see Fig.\ref{fig: Fig. 1}a), considering $2$-simplex interactions along with $1$-simplex interactions represented using various colors (see Fig.\ref{fig: Fig. 1}b). The dynamical equation for each oscillator in the network is given by:

\begin{align}
\dot{\theta_i} &= \omega_i 
+ \frac{k_1}{N}\sum_{j=1}^{N} 
    \sin(\theta_j - \theta_i - \beta) \nonumber\\
&\quad
+ \frac{k_2}{N^2}\sum_{j=1}^{N} \sum_{l=1}^{N} 
    \sin(2\theta_j - \theta_l - \theta_i - \beta)
+ \zeta_i(t),
\label{Eq. 1}
\end{align}

Here, $\omega_i$ and $\theta_i$ denote the intrinsic natural frequency and phase of the oscillator, $i$ ($i=1,2,\ldots, N$). Here, $N$ is the total number of oscillators or nodes in the network, and $\beta$ is the phase-lag parameter that introduces \emph{frustration} into the system. We consider nonidentical oscillators by taking distinct natural frequencies $\omega_i$ for each node, drawn from the Lorentzian distribution\cite{PhysRevE.98.022207}, $g(\omega) = \frac{\Delta}{\pi\left[(\omega - \omega_0)^2 + \Delta^2\right]}$, where $\omega_0$ is the mean frequency and $\Delta$ is the width of the distribution. Throughout this study, we analyze the nonzero-frequency case, i.e., $\omega_i \neq 0$.  

The second term in Eq.~\ref{Eq. 1} corresponds to pairwise ($1$-simplex) interactions with coupling strength $k_1$, while the third term accounts for higher-order ($2$-simplex) interactions with coupling strength $k_2$. The last term introduces stochasticity in the system, where $\zeta_{i}(t)$ represents Gaussian white noise with zero mean, $\delta$ correlated in time, and independent across oscillators. Specifically, 
$\langle \zeta_i(t) \rangle = 0$ and $\langle \zeta_i(t)\,\zeta_j(t') \rangle = 2D\,\delta_{ij}\,\delta(t-t')$,  where $D \geq 0$ denotes the noise strength.  

In the Kuramoto framework, the degree of synchronization is quantified by the magnitude $r$ of the complex order parameter $z$. It is often referred to as the order parameter, and it takes values in the range $0 \leq r \leq 1$. The general expression for $z$ is given by \cite{CARBALLOSA2023114197}:

\begin{equation}
  z_m(t) = r \, {e^{im\psi_m}} =\frac{1}{N} \, \sum_{j=1}^{N} e^{i m \theta_j}
  \label{Eq. 2}
\end{equation}

where $m$ represents the order of interactions; this implies $m=1$ for $1$-simplex and $m=2$ for $2$-simplex interactions, respectively. Similarly, $\psi_m$ denotes the mean phase, while $\theta_j$ represents the phase of the $j$-th oscillator.
The order-parameter equation for the pairwise interactions is written as;
\begin{equation}
    z_1(t) = r_1 \, {e^{i\psi_1}} = \frac{1}{N} \, \sum_{j=1}^{N} e^{i \theta_j} 
    \label{Eq. 3}
\end{equation}
where $r_1 = |z_1|$ is the amplitude of the order-parameter $z_1$.

Similarly, by setting $m=2$ in the general order parameter definition (Eq.~\ref{Eq. 2}), one obtains a higher-order order parameter that quantifies the degree of synchronization due to triadic ($2$-simplex) interactions in the system, given by
\begin{equation}
  z_2(t) = r_2 \, {e^{2i\psi_2}} =\frac{1}{N} \, \sum_{j=1}^{N} e^{i 2 \theta_j}\label{Eq. 4}
\end{equation}
where $r_2 = |z_2|$ is the amplitude of the complex order-parameter $z_2$. 
% $r_1$ and $r_2$ measure the level of synchronization in the network. 

%  $R \, (=r_1)$ is often referred to as the order parameter, and it takes values in the range $0 \leq R \leq 1$. 
Throughout this paper, we take $R(= r_1)$ as a measure of the system's synchronization. The case $R=0$ corresponds to a completely desynchronized state, where all oscillators are uniformly spread around the unit circle, whereas $R=1$ indicates a fully synchronized state with all oscillators perfectly phase-aligned (a one-cluster state).

%\section{\label{Sec:level3}Results}
The study begins with a numerical investigation of the synchronization dynamics of a globally coupled network of nonidentical SK oscillators, modeled by Eq.~\ref{Eq. 1} with a network of size $N=100$. The initial conditions for the system are an equally populated bicluster state with $50$ oscillators having phase $\phi=0$, and the remaining oscillators have phase $\phi=\pi$. The intrinsic frequencies $\omega$ of the oscillators are drawn from a Lorentzian frequency distribution with a mean frequency $\omega_0 = 0$ and width $\Delta=0.5$. To find the influence of stochasticity on the system's behavior, three distinct noise strengths are considered: $D=0.0, 0.5, 1.0$.\\

%\subsubsection{\label{Sec:level2}Synchronization Dynamics in the $(k_1, \beta)$ Space}
\textbf{Synchronization in the $(k_1, \beta)$ Space:}
To analyze the emergence of synchronization in the coupled system modeled by Eq. \ref{Eq. 1}, we first examine the order parameter in the $(k_1, \beta)$ parameter space. Specifically, $k_1$ and $\beta$ are varied within the ranges $k_1 \in [-15, 15]$ and $\beta \in [-\pi, \pi]$, respectively. For each fixed value of $k_2$, we investigate synchronization among oscillators for three representative noise strengths: $D = 0.0$, $0.5$, and $1.0$ by calculating the order parameter $R$.
Since the range of $1$-simplex interaction strength $k_1 \in [-15, 15]$, we choose five distinct values of $2$-simplex interaction strength, $k_2 = -20, -10, 0, 10$ and $20$, thereby covering the transition from strongly repulsive to strongly attractive triadic interactions in the network.

\begin{figure*}
 \centering
    \includegraphics[width=1.0\textwidth, height=0.85\textheight]{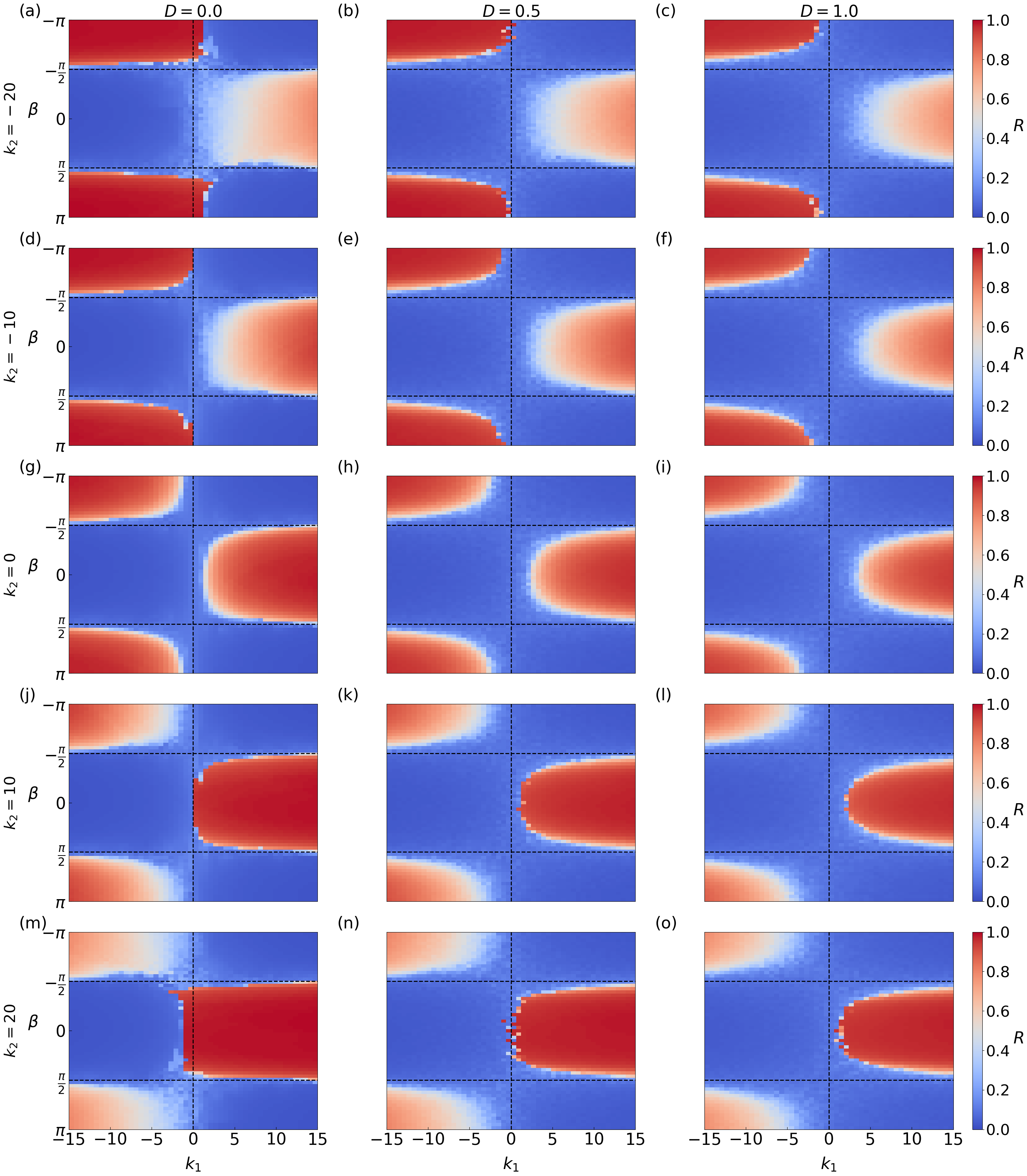}
    \caption{Synchronization landscapes in the $(k_1, \beta)$ parameter space for a network of $N = 100$ frustrated stochastic Kuramoto oscillators incorporating higher-order ($2$-simplex) interactions. The degree of collective coherence is measured by the order parameter $R$. The color map encodes the value of $R$, with dark blue ($R = 0$) corresponding to a fully desynchronized state and red ($R = 1$) indicating complete synchronization. Each column represents a distinct noise strength, $D = 0$, $0.5$, and $1$, while each row corresponds to a fixed value of the $2$-simplex coupling strength, $k_2 \in {-20, -10, 0, 10, 20}$. Specifically, panels $(a$–$c)$ depict $k_2 = -20$, $(d$–$f)$ $k_2 = -10$, $(g$–$i)$ $k_2 = 0$, $(j$–$l)$ $k_2 = 10$, and $(m$–$o)$ $k_2 = 20$. The figure highlights how varying higher-order coupling and noise strength jointly modulate the onset and extent of synchronization within the network of coupled oscillators. 
    }\label{fig: Fig. 2}
\end{figure*}

Figure~\ref{fig: Fig. 2} illustrates the dynamical regions in the $(k_1, \beta)$ parameter space, and several salient trends can be identified that influence the emergence of synchronization within the network. For example, for the noiseless case ($D = 0$), as shown in Figs.~\ref{fig: Fig. 2}(a,d,g,j,m), the network exhibits synchronized and desynchronized regimes in ($k_1, \beta$) space, and the extent of these regions strongly depends on the $2$-simplex strength $k_2$. 
For example, Fig.~\ref{fig: Fig. 2}(g), which corresponds to the absence of higher-order interactions ($k_2 = 0$), shows a synchronization region that is symmetrically distributed around $\beta = 0$. The extent of the synchronized and desynchronized regions in the ($k_1, \beta$) space depends on the values of  $k_2$. Specifically, for $k_1 > 0$, synchronization emerges when $\beta \in (-\pi/2, \pi/2)$, whereas for $k_1 < 0$, synchronization occurs in the regions $\beta \in (\pi/2, \pi]$ and $\beta \in (-\pi/2, -\pi]$. This indicates that positive (mutual) pairwise couplings favor synchronization when the intrinsic phase frustration of the oscillators is relatively small.
%extends over both high positive and negative values of $k_1$, indicating that if nodes are pair-wise interactions are mutual then, to attain synchronization, internal phase or frustration among the can emerge for either sign of the $k_1$, provided that $\beta$ lies within favorable intervals—specifically, 

At strong negative $2$-simplex coupling strength ($k_2 = -20$ cf. Fig.~\ref{fig: Fig. 2}(a)), synchronization predominantly appears when the intrinsic phase lag lies within $\beta \in (-\pi/2, -\pi] $ and $\beta \in (\pi/2, \pi]$, and $k_1 \in [-15, 1.5)$. This indicates that for strong repulsive higher-order interactions, synchronization is sustained only when pairwise interactions are repulsive or weakly mutual.

As $k_2$ decreases to $-10$ (cf. Fig.~\ref{fig: Fig. 2}(d)), synchronized regimes emerge for negative values of $k_1$, particularly when the phase lag lies within the ranges $\beta \in (-\pi/2, -\pi]$ and $\beta \in (\pi/2, \pi]$. For large negative values of $k_1$, the synchronized regions expand and cover nearly the entire aforementioned $\beta$ range, compared to the narrower bands observed at less negative $k_1$. Additionally, we observe that synchronization gradually begins to appear on the right-hand side of the $(k_1, \beta)$ plane for larger positive $k_1$ values with increasing $k_2$ values. This behavior suggests that weaker repulsive triadic couplings can partially alleviate phase frustration and facilitate the emergence of coherence among oscillators, even when $k_1$ is positive and large.

For $k_2 > 0$, the synchronization regimes shift markedly in $(k_1, \beta)$ space; for instance, at $k_2 = 10$ (cf Fig.~\ref{fig: Fig. 2}(j)), a distinct synchronized region (red lobe) emerges for $k_1 \in [0, 15]$, and weak synchronization begins to appear even for slightly negative $k_1$, reflecting the partial stabilization effect of positive higher-order interactions. This tendency becomes more pronounced at $k_2 = 20$ (cf Fig.~\ref{fig: Fig. 2}(m)), where the synchronized region broadens and forms an almost symmetric red domain around $\beta = 0$, extending across $\beta \in (-\pi/2, \pi/2)$. This shift demonstrates that strong positive triadic interactions can effectively counteract the desynchronization influence of phase frustration and weaker pairwise coupling, thereby restoring coherence in the network.

These observations collectively suggest that when the intrinsic phase lag lies within $\beta \in (-\pi/2, \pi/2)$, oscillators either \emph{lag} behind ($\beta > 0$) or \emph{lead} ($\beta < 0$) their neighbors, resulting in weak phase frustration among the nodes. To overcome this weak frustration and sustain synchronization, $k_1$ must be sufficiently large, although its critical value depends sensitively on $k_2$. For example, from Fig.~\ref{fig: Fig. 2}(m), it is evident that when $k_2 = 20$, synchronization emerges from $k_1 \approx -1.5$, whereas for $k_2 = -10$, synchronization appears only for larger $k_1$ values near $k_1 \approx 14$ (see Figs.~\ref{fig: Fig. 2}(d)). 

In contrast, when the intrinsic phase lag lies within $\beta \in (\pm \pi/2, \pm \pi]$, the degree of frustration is stronger, leading to more repulsive coupling among oscillators. In such cases, synchronization can emerge only when $k_2$ is also repulsive (negative) and $k_1$ is predominantly negative, although small positive $k_1$ values may still permit weak coherence for strongly negative $k_2$ (see fig. ~\ref{fig: Fig. 2}(a)). Hence, synchronization in this region is possible only when the coupling strengths are sufficiently negative to counterbalance the strong intrinsic phase frustration effect on synchronization.

The influence of stochasticity on these patterns is captured in the middle and right columns of Fig.~\ref{fig: Fig. 2}, corresponding to $D = 0.5$ and $D = 1.0$, respectively. From Figs.~\ref{fig: Fig. 2}(b, e, h, k, n) and Figs.~\ref{fig: Fig. 2}(c, f, i, l, o), it is evident that introducing noise progressively diminishes the synchronized regimes across the same $(k_1, \beta)$ parameter range. We also observe that the transition from desynchronization to synchronization becomes smoother in the presence of noise. This demonstrates that stochastic fluctuations disrupt phase alignment among oscillators, transforming the sharp first-order synchronization transitions observed in the deterministic system into continuous and second-order transitions at higher noise intensities.

Overall, these observations highlight that synchronization in a network of coupled oscillators arises from the intricate interplay among pairwise and triadic interactions, intrinsic phase frustration, and stochastic fluctuations. When intrinsic phase frustration is weak, broadly, synchronization is favoured for both positive coupling strengths. However, under strong phase frustration, coherence emerges only when both coupling strengths are repulsive. High levels of stochasticity tend to suppress overall coherence among oscillators and simultaneously smooth the transition from desynchronization to synchronization.

\begin{figure*}
 \centering
    \includegraphics[width=1.0\textwidth, height=0.6\textheight]{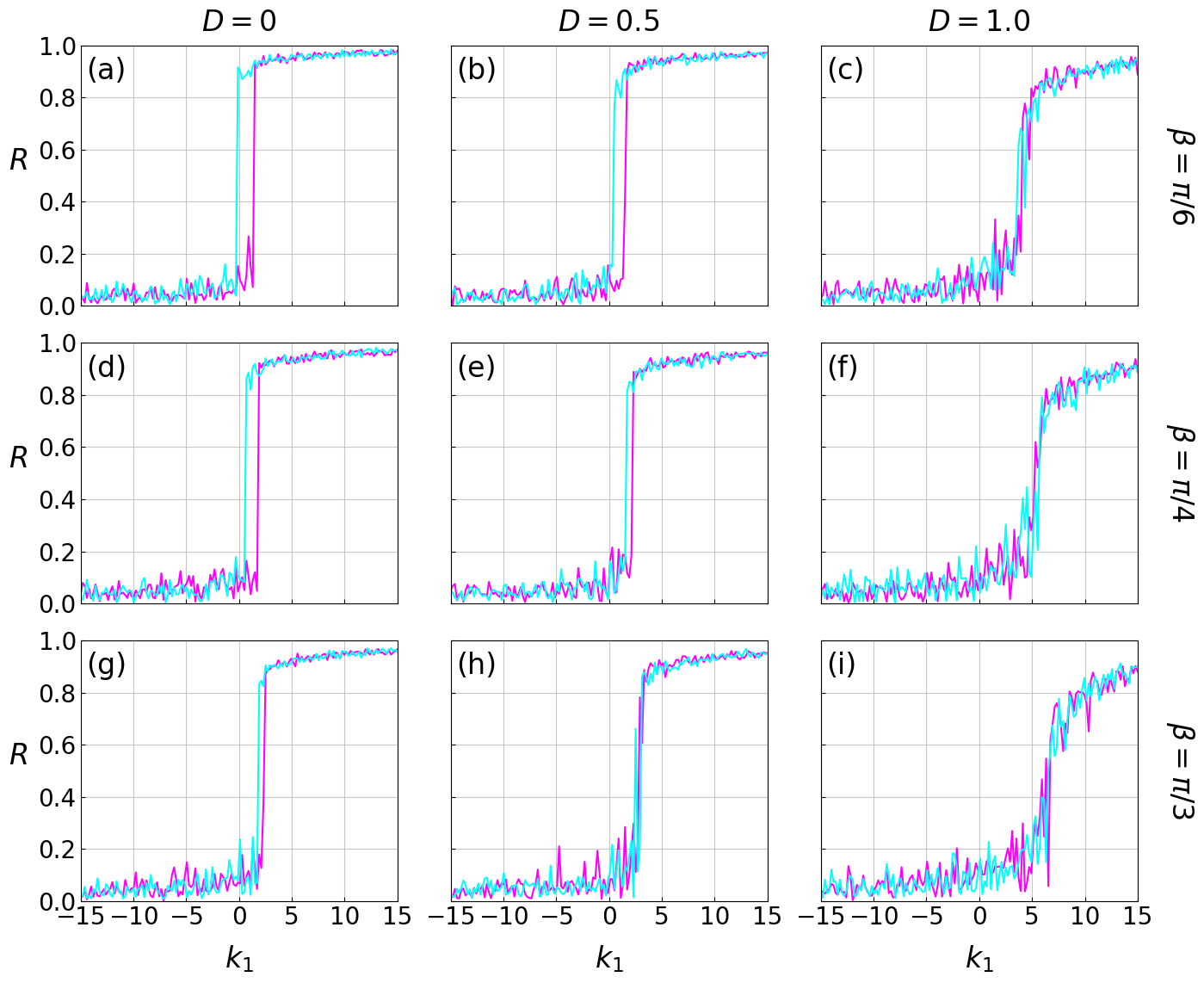}
    \caption{Variation of the order parameter $R$ as a function of the pairwise coupling strength $k_{1}$ 
    for forward-sweep (magenta color) and backward-sweep (cyan color) simulations, at a fixed higher-order interaction strength $k_{2}=10$ for a system of $N=100$ oscillators. The columns correspond to different noise strengths: panels $(a,d,g)$, $(b,e,h)$, and $(c,f,i)$ represent $D=0$, $0.5$, and $1.0$, respectively, while the rows correspond to different phase-lag values: panels $(a$--$c)$, $(d$--$f)$, and $(g$--$i)$ correspond to $\beta=\pi/6$, $\pi/4$, and $\pi/3$, respectively. The natural frequencies $\omega_i$ are drawn from a Lorentzian distribution with $\omega_0=0$ and width $\Delta=0.5$. The initial condition is an equally populated bi-cluster state in which half of the oscillators have phase $\phi=0$ and the remaining half of the oscillators have phase $\phi=\pi$. The figure demonstrates that the width of the hysteresis loop decreases systematically with increasing phase lag $\beta$ and noise strength $D$.} \label{fig: Fig. 3}
\end{figure*}

\begin{figure*}
 \centering
    \includegraphics[width=1.0\textwidth, height=0.6\textheight]{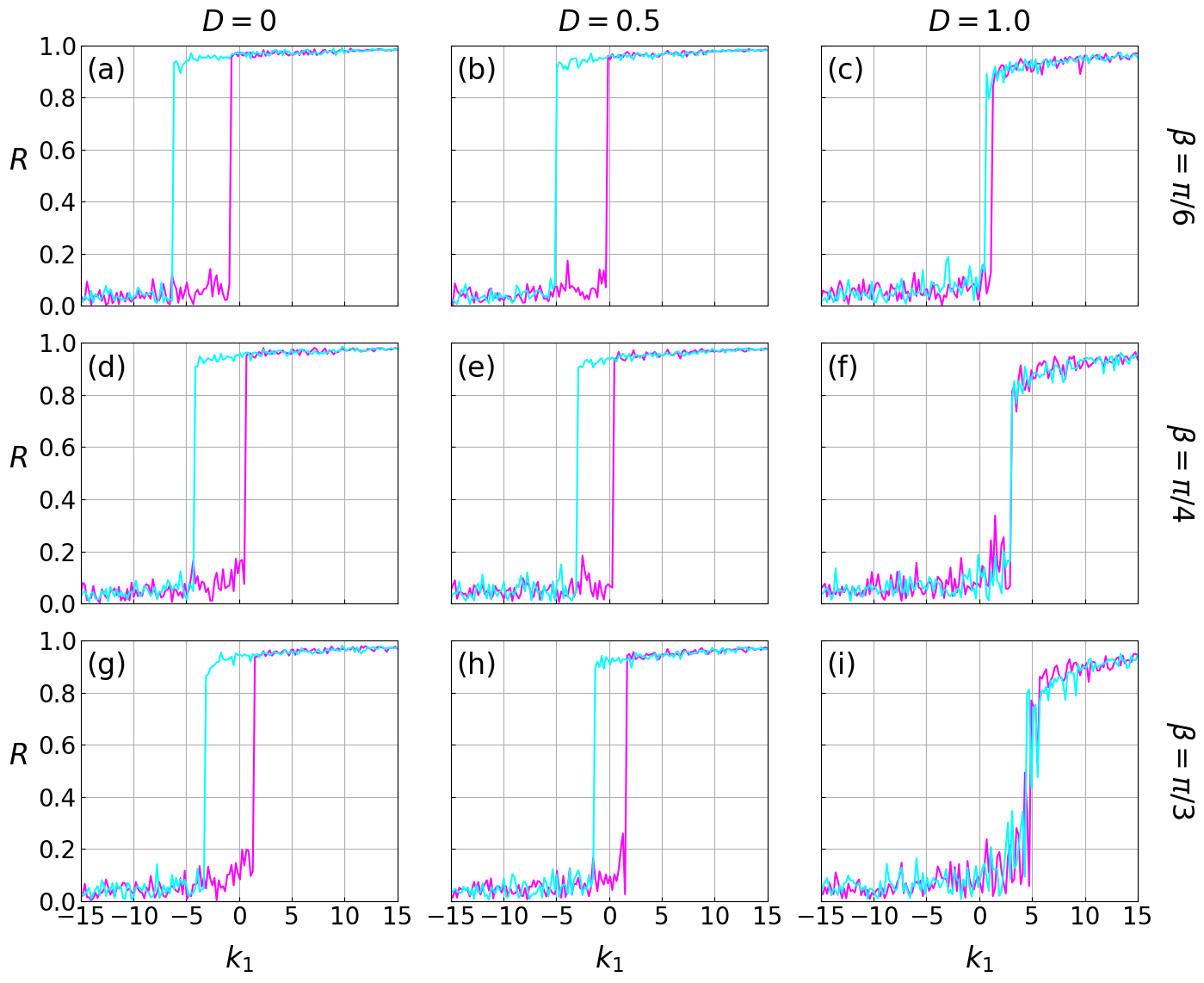}
    \caption{Variation of the order parameter $R$ as a function of the pairwise coupling strength $k_{1}$ 
    for forward sweep (magenta color) and backward sweep (cyan color) simulations, at a fixed higher-order interaction strength $k_{2}=20$ for a system of $N=100$ oscillators. Each column corresponds to a different noise strength: panels $(a,d,g)$ correspond to $D=0$, panels $(b,e,h)$ are for the case when $D=0.5$, and panels $(c,f,i)$ correspond to $D=1.0$; each row corresponds to a different phase-lag value: 
    panels $(a-c)$ for  $\beta=\pi/6$, panels $(d-f)$ for $\beta=\pi/4$, and panels $(g-i)$ for $\beta=\pi/3$. The natural frequencies $\omega_i$ are drawn from a Lorentzian distribution with $\omega_0=0$ and a width $\Delta=0.5$. The initial condition is an equally populated bi-cluster state in which half of the oscillators have phase $\phi=0$, and the remaining half of the oscillators have phase $\phi=\pi$. The figure demonstrates that the width of the hysteresis loop decreases systematically with increasing phase lag $\beta$ and noise strength $D$.} \label{fig: Fig. 4}
\end{figure*}

\textbf{Forward and backward simulations:}
In the forward simulation, the system is initialized in an equally populated bicluster configuration and evolves while the value of $k_1$ is gradually increased. Once the system reaches a synchronized state, this final configuration is used as the initial condition for the backward simulation, in which $k_1$ is decreased in small, gradual steps (refer to Fig.~\ref{fig: Fig. 3} and Fig.~\ref{fig: Fig. 4}).

The forward sweep, FS (magenta), and backward sweep, BS (cyan), branches reveal that the coupled system may exhibit hysteresis, and the width of this hysteresis depends on $\beta$, $D$, and $k_2$. For example, Figs.~\ref{fig: Fig. 3}(a,d,g) show clear hysteresis loops whose widths progressively shrink as the phase lag $\beta$ increases, for a fixed set of the remaining parameters. Similarly, Figs.~\ref{fig: Fig. 3}(a,b,c) illustrate that hysteresis persists only up to a certain level of stochasticity, and the hysteresis width decreases as the noise intensity $D$ increases.

From Fig.~\ref{fig: Fig. 3} and Fig.~\ref{fig: Fig. 4}, it is evident that the higher-order interaction strength $k_2$ plays a major role in widening the bistable region. This behavior differs from the finite-size effect reported in \cite{suman2024finite}, where the system size controls the width of the bistable region.

To examine the robustness of the synchronization transition with respect to the initial conditions, we performed additional numerical simulations using two different phase distributions, namely, random phases uniformly distributed in $[0,2\pi]$ and phases uniformly distributed in $[0,\pi/2]$, in addition to the equally populated anti-phase bi-cluster state. As shown in Fig.~\ref{fig: Fig. 6}, all three initial conditions exhibit qualitatively similar synchronization transitions and hysteresis behavior. Although the transition points show minor quantitative differences, the overall synchronization behavior is determined primarily by the model parameters rather than the choice of initial conditions. To understand the mechanism behind the emergence of bistability between synchronized and desynchronized states in our coupled network, we further analyze the system dynamics theoretically, as described in the following sections.

\section{\label{sec:level3}Low-Dimensional Model and Analytical Results}

Fig.~\ref{fig: Fig. 2} illustrates that the system undergoes different types of transitions from desynchronized to synchronized states, depending on the parameter values. For instance, in Fig.~\ref{fig: Fig. 2}(j), when the phase lag $\beta$ approaches zero, even a small increase in the pairwise coupling strength $k_1$ near $k_1 = 0$ (in the positive direction) leads to an abrupt jump to synchronization, indicating a first-order transition. In contrast, some transitions occur more smoothly. For example, Fig.~\ref{fig: Fig. 2}(f) shows a gradual increase in coherence with increasing $k_1$, characteristic of a second-order phase transition. Fig.~\ref{fig: Fig. 3} and Fig.~\ref{fig: Fig. 4} clearly exhibit both first- and second-order transitions, depending on the specific combination of parameters $(k_2, \beta, D)$.

\begin{figure*}
 \centering
    \includegraphics[width=1.0\textwidth, height=0.6\textheight]{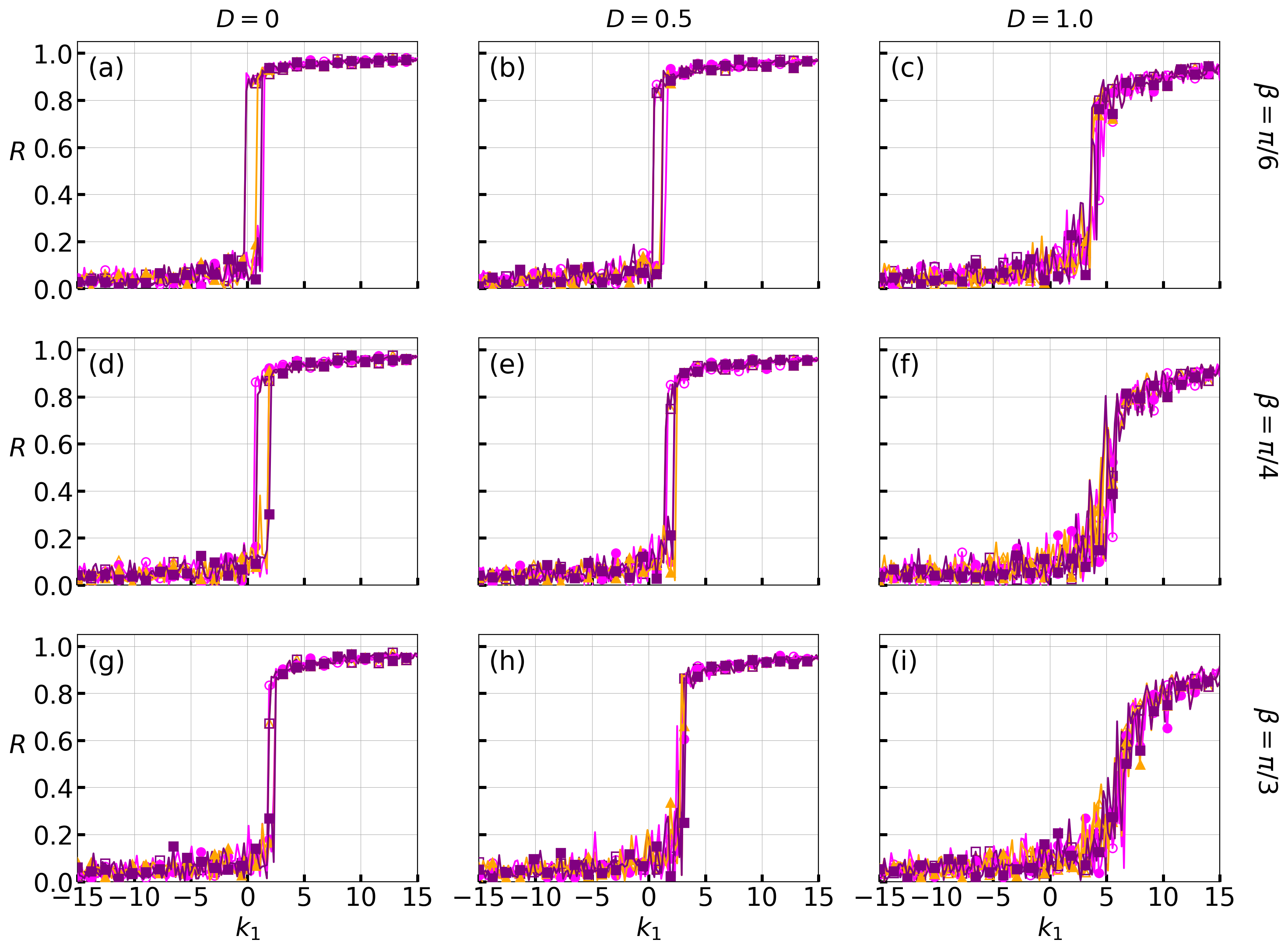}
    \caption{Variation of the synchronization order parameter $R$ as a function of the pairwise coupling strength $k_1$ for $N=100$ globally coupled frustrated stochastic Kuramoto-Sakaguchi oscillators with fixed $2$-simplex interaction strength $k_2=10$. Simulations are performed for three initial conditions: (i) an equally populated anti-phase bi-cluster state (magenta circles), (ii) random phases uniformly distributed in $[0,2\pi]$ (orange triangles), and (iii) phases uniformly distributed in $[0,\pi/2]$ (purple squares). Hollow and filled symbols of the same color denote the forward and backward sweeps, respectively. The columns correspond to different noise strengths: panels $(a,d,g)$, $(b,e,h)$, and $(c,f,i)$ represent $D=0$, $0.5$, and $1.0$, respectively, while the rows correspond to different phase-lag values: panels $(a$--$c)$, $(d$--$f)$, and $(g$--$i)$ correspond to $\beta=\pi/6$, $\pi/4$, and $\pi/3$, respectively. The synchronization transition and hysteresis remain qualitatively unchanged for all three initial conditions, demonstrating the robustness of the observed behavior.} 
    \label{fig: Fig. 5}
\end{figure*}

\begin{figure*}
 \centering
    \includegraphics[width=1.0\textwidth, height=0.6\textheight]{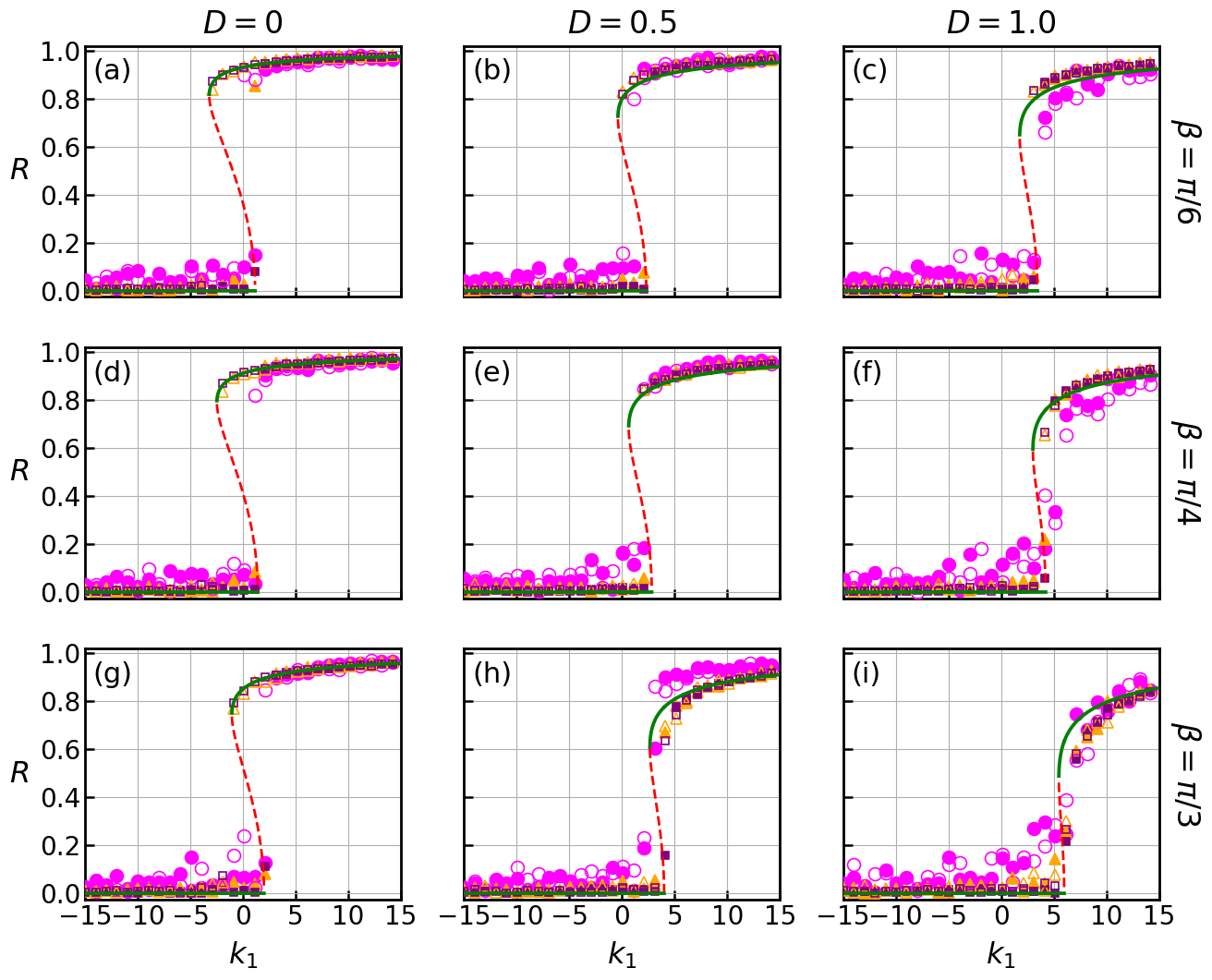}
    \caption{Variation of the synchronization order parameter $R$ as a function of the pairwise coupling strength $k_1$. Symbols denote the numerically obtained Kuramoto order parameter given by Eq. \ref{Eq. 3} for network sizes $N=100$, $1000$, and $5000$, while the solid green and dotted red curves represent the stable and unstable theoretical solutions, respectively, obtained from Eq.~\ref{Eq. 13}. Each panel of the plot showcases the comparison of numerical simulations with the analytical solutions obtained via solving the reduced OA equation Eq. \ref{Eq. 13}. The forward simulation curve for $N=100$ is shown via filled magenta circles, for $N=1000$, filled orange triangles, and for $N=5000$, filled purple squares. Similarly, for $N=100$, hollow magenta circles, for $N=1000$, hollow orange triangles, and for $N=5000$, hollow purple squares represent the backward simulation curves. The green solid line represents the stable branch, whereas the dashed red line represents the unstable branch of the solution of the reduced OA equation. Panels \emph{a, d, g} corresponds to $D=0$, panels \emph{b, e, h} corresponds to $D=0.5$, and panels \emph{c, f, i} corresponds to $D=1.0$. Similarly, panels \emph{a, b, c} is for $\beta=\pi/6$, panels \emph{d, e, f} is for $\beta=\pi/4$, and panels \emph{g, h, i} is for $\beta=\pi/3$.}
    \label{fig: Fig. 6}
\end{figure*}

To capture such transitions analytically, we employ the Ott-Antonsen (OA) approach \cite{ott2008low}, which reduces the continuum dynamics of coupled phase oscillators to a low-dimensional system on an invariant manifold in the thermodynamic limit ($N \rightarrow \infty$). Underlying this approach, the system is characterized by a density function $\rho(\theta, \omega, t)$, and its temporal evolution is governed by the Fokker-Planck (continuity) equation \cite{Risken1989}.

For the stochastic dynamical model given by Eq.~\ref{Eq. 1}, the corresponding Fokker-Planck equation is \cite{marui2025synchronization}

\begin{align}
\frac{\partial \rho}{\partial t}
&= -\,\frac{\partial (v\rho)}{\partial \theta}
   + D\,\frac{\partial^2 \rho}{\partial \theta^2}
\label{Eq. 5}
\end{align} 

Here, $D$ denotes the noise strength and $v$ represents velocity, which is defined as (please refer to Sec. 1 of the supplementary material for detailed steps to find $v$); 

\begin{equation}
    v(\theta, \omega, t) = \omega + \frac{1}{2i}{\left(He^{-i(\theta+\beta)} - H^{*}e^{i(\theta+\beta)}\right)} \label{Eq. 6}
\end{equation}

where, $H = (k_1z_1 + k_2z_2z_1^{*})$ and $H^{*}$ is the complex conjugate of $H$. The order parameter ($z_m = r_m e^{i\psi_m}, m =1, 2$) in the continuum limit $(N \rightarrow\infty)$ can be written as:

\begin{align}
z_1 &= r_{1} e^{i\psi_1} %\nonumber\\
    = \int_{0}^{2\pi} \!\!\int_{-\infty}^{\infty} e^{i\theta}\, \rho(\theta, \omega, t) d\omega d\theta
\label{Eq. 7}
\end{align}

\begin{align}
     z_2 &= r_{2}e^{i\psi_2} %\nonumber\\
     = \int_{0}^{2\pi} \int_{-\infty}^{\infty} e^{2i\theta}\rho(\theta, \omega, t) d\omega d\theta
     \label{Eq. 8}
\end{align}

where $z_1$ corresponds to the order parameter for pairwise interaction and $z_2$ represents the order parameter for $2-simplices$ interactions. The solution of Eq. \ref{Eq. 5}, using the Ott-Antonsen ansatz, can be written by expanding the probability density function ($\rho$), in terms of the Fourier series:

\begin{equation}
    \rho(\theta, \omega, t) = \frac{g(\omega)}{2\pi}\left(1 + \sum\limits_{n = 1}^{\infty} \rho_n(\omega, t)e^{in\theta} +  c.c\right) 
    \label{Eq. 9}
\end{equation}

Here, $g(\omega)$ is the Lorentzian frequency distribution, and $\rho_n(\omega, t)$ is the coefficient of the $n$-th term of the Fourier series, and \emph{c.c} is the complex conjugate.

Substituting Eqs. ~\ref{Eq. 6} and \ref{Eq. 9} into Eq. ~\ref{Eq. 5}, and equating the coefficients of equal Fourier harmonics yields an infinite hierarchy of evolution equations for the Fourier coefficients $\rho_n$ (see Eq.~S16 in Sec.~1A of the Supplemental Material).
In the presence of Gaussian white noise, the OA manifold is no longer exactly invariant because the diffusion term in the Fokker-Planck equation generates an infinite hierarchy of Fourier modes. Although more accurate low-dimensional reductions, such as the Gaussian ansatz and the circular-cumulant approach, have been developed for noisy oscillator populations \cite{PhysRevLett.120.264101,10.1063/1.5053576}, the leading-order OA closure employed here provides an analytically tractable approximation that captures the bifurcation structure and synchronization transitions observed over the parameter range considered. Previous studies have demonstrated that this approximation remains accurate for weak to moderate noise strengths \cite{10.1063/1.5053576}.

Applying the OA closure to Eq. ~\ref{Eq. 1} yields the following approximate evolution equation for the first Fourier mode (see Sec.~S1A of the Supplemental Material for the detailed derivation)
 
\begin{equation}
    \dot{\alpha}(\omega, t) = -(i\omega + D)\alpha(\omega, t) + \frac{1}{2}\left(H^*e^{i\beta} - He^{-i\beta}\alpha^2(\omega,t)\right) 
    \label{Eq. 10}
\end{equation}

For a Lorentzian natural frequency distribution, the Ott--Antonsen function $\alpha(\omega,t)$ is assumed to be analytic in the lower half of the complex $\omega$-plane. Consequently, the contour integral is evaluated using the residue theorem at the pole $\omega=\omega_0-i\Delta$. Identifying the residue with the complex order parameter through $z_1=\alpha^{*}(\omega_0-i\Delta,t)$, we obtain the following approximate reduced equation:
\begin{equation}
    \dot{z_1} = (i \omega_0 - (\Delta + D))z_1 + h - h^{*}z_1^{2}
    \label{Eq. 11}
\end{equation}
where $h=\frac{1}{2}He^{-i\beta}$ and (see Sec. 1B in the supplementary file for the detailed derivation).

Since the OA closure is approximate for $D>0$, the corresponding relation between the first and second order parameters is written as $z_2\simeq z_1^2$, substituting the resulting mean field $H = \left(
k_1+k_2|z_1|^2
\right)z_1$ into the reduced order-parameter Eq.\ref{Eq. 11}, we obtain 
\begin{align}
\dot{z}_1
={}&
\left(
i\omega_0-(\Delta+D)
\right)z_1
% \nonumber\\
% &
+
\frac{1}{2}
\left(
k_1+k_2|z_1|^2
\right)
\left(
e^{-i\beta}
-
|z_1|^2e^{i\beta}
\right)z_1.
\label{eq: Eq. 12}
\end{align} 

Further expressing the complex order parameter in polar form, $z_1=re^{i\psi}$, we obtain the following approximate amplitude equation (see Sec.~1B of the Supplemental Material for the detailed derivation):

\begin{equation}
    \dot{r} = -(D + \Delta)r + \frac{\cos \beta}{2} \left(k_1 r + k_2 r^3\right)\left(1 - r^2\right) 
    \label{Eq. 13}
\end{equation}

Eq. \ref{Eq. 13} provides a reduced low-dimensional description of the collective dynamics. By analyzing its steady-state solutions and their linear stability, we determine the critical coupling strengths, identify the nature of the synchronization transition and the associated bifurcation structure, and quantify the extent of the bistable region. In the absence of noise ($D=0$), Eq. \ref{Eq. 13} reduces to the analytical result reported in Ref.~\cite{PhysRevE.108.034208}, thereby confirming the consistency of the present formulation with the existing theory.

\begin{figure*}
 \centering
    \includegraphics[width=1.0\textwidth, height=0.25\textheight]{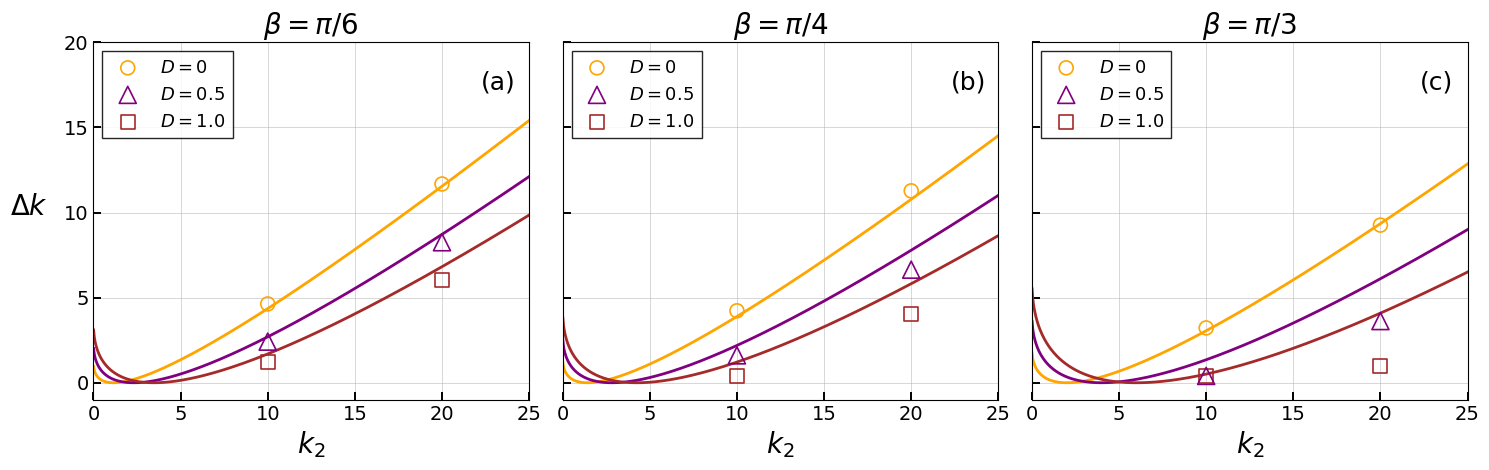}
    \caption{Variation of the hysteresis width ($\Delta k$) as a function of the $2$-simplex interaction strength ($k_2$) for different noise strengths ($D$) and three values of the phase lag, $\beta=\pi/6$, $\beta=\pi/4$, and $\beta=\pi/3$, shown in panels (a), (b), and (c), respectively. The solid curves represent the analytical prediction computed from $\Delta k = |k_{1c}-k_{1}^{SN}|$, while the hollow circles, triangles, and squares denote the numerically obtained hysteresis width for a network of size $N=5000$ at three different values of $D$ for each $\beta$ mentioned in the figure, respectively. For noise-less case ($D=0$), the numerical results agree with the analytical prediction (see orange curves and circles). As the noise strength increases, the observed hysteresis width decreases due to noise-induced switching. Consequently, the deviation between the numerical and analytical results becomes more pronounced with increasing $D$ and $k_2$. The critical value of higher-order interaction varies as $k=2(\Delta+ D)/\cos\beta$}. %The system's initial behavior exhibits a larger hysteresis width. With increasing value of $k_2$ the hysteresis width increases and overtakes after $k_2>2$. The variations in the hysteresis width is evident from Fig.~\ref{fig: Fig. 3} and Fig.~\ref{fig: Fig. 4}.
  \label{fig: Fig. 7}
\end{figure*}

To examine finite-size effects and assess the validity of the OA reduction, we performed numerical simulations for networks of sizes $N=100$, $1000$, and $5000$, and compared the results with the theoretical predictions obtained by solving the low-dimensional order-parameter equation, Eq.~\ref{Eq. 13}. As shown in Fig.~\ref{fig: Fig. 7}, the agreement between the numerical order parameter (Eq.~\ref{Eq. 3}) and the theoretical solutions (solid green and dotted red curves denote the stable and unstable solutions, respectively) improves systematically with increasing network size. This behavior is expected because the OA reduction is derived in the thermodynamic limit ($N\rightarrow\infty$), while finite-size fluctuations, which scale as $O(N^{-1/2})$, become progressively weaker as $N$ increases, causing the finite-$N$ dynamics to approach the mean-field prediction. Furthermore, the agreement between theory and simulations is excellent in the deterministic limit ($D=0$). As the noise strength increases, however, the agreement gradually deteriorates because the OA reduction provides an approximate description of the stochastic dynamics. Nevertheless, it successfully captures the qualitative synchronization behavior and the underlying bifurcation structure over the considered parameter range.

To further understand the bistability phenomenon in the network, we analyze the reduced dynamics (Eq. \ref{Eq. 13}), identifying the fixed points and their stability, as well as bifurcations.

%\section{Analytical Results}
\textbf{Fixed points}:
The solutions of Eq. \ref{Eq. 13} are obtained by setting $\dot{r} = 0$. One of the trivial solutions is ${r^* = 0}$, while the other solutions are obtained by solving Eq. ~\ref{Eq. 15}. That is, 

\begin{equation} 
    -(\Delta + D) + \frac{\cos \beta}{2} \left(k_1  + k_2r^2\right)\left(1 - r^2\right) = 0 \label{Eq. 14}
\end{equation}

 We obtain the following four nontrivial fixed points:

\begin{equation}
    r^*_{\pm} = \pm{ \sqrt{\frac{(k_2 - k_1) \pm\sqrt{(k_2 - k_1)^2 +4k_2(k_1 - A)}}{2k_2}}} {\label{Eq. 15}}
\end{equation}

where $A = 2(\Delta+D)/\cos\beta$. The detailed steps for finding these fixed points are given in the supplementary information material (please refer to Sec. 2 of the supplementary material).

Since $r \in [0,1]$, the only positive solution is physically relevant. Moreover, as we seek positive real solutions, the discriminant of Eq. \ref{Eq. 15} must be non-negative.
Thus, we find only two physically relevant non-zero fixed points, which are the following:

\begin{equation}
  r^* = \sqrt{\frac{(k_2 - k_1) \pm\sqrt{(k_2 - k_1)^2 +4k_2(k_1 - A)}}{2k_2}}
  \label{Eq. 16}
\end{equation}

\textbf{Stability of fixed points}:
From the linear stability analysis, we find that $r^{*} = 0$ is always stable for $\cos\beta =0$ (and $\cos\beta >0$). Similarly, it will be stable if 
$k_1 < \frac{2(\Delta + D)}{\cos\beta}$ (and $\cos\beta <0$). And it will be stable if $k_1 > \frac{2(\Delta + D)}{\cos\beta}$.

Linear stability analysis of the two physically relevant fixed points, the larger and small-amplitude solutions (see Eq. \ref{Eq. 15}), shows that the large-amplitude fixed point $r^{*}$ is stable, whereas the small-amplitude fixed point is unstable.

\textbf{Bifurcation}: From Eq.~\ref{Eq. 13}, the reduced amplitude equation may take the normal form of either a subcritical or a supercritical pitchfork bifurcation, depending on the signs of the coefficients of $r^{3}$ and $r^{5}$ (please refer to the supplementary Sec. 2 for the complete derivation). The figures, S1-S6 (please refer to the supplementary material, Sec. 4), present a clear picture of subcritical and supercritical pitchfork bifurcations, depending on the parameter values. Further, we find critical points of the bifurcations. When the parameter $k_1$ is varied in the forward direction, the bifurcation point at which the trivial solution $r^{*}=0$ loses stability is $k_{1c} = \frac{2(\Delta + D)}{\cos\beta}$. This transition marks the loss of stability of the trivial state and the emergence of the stable nontrivial state. In the backward direction, the two nontrivial fixed points (one stable and one unstable) collide at $r^{*} = \sqrt{\frac{k_2 - k_1}{2k_2}},$ indicating a saddle-node bifurcation. The corresponding critical value is $k_{1}^{\mathrm{SN}} = -\,k_2 + 2\sqrt{\frac{2k_2(\Delta + D)}{\cos\beta}}$.

Fig.~\ref{fig: Fig. 7} shows theoretical calculated hysteresis width, $\Delta k = |k_{1c}-k_{1}^{\mathrm{SN}}|$ as a function of $k_2$ for $D=0$, $0.5$, and $1$, and $\beta=\pi/6$, $\pi/4$, and $\pi/3$. The analytical prediction shows that $\Delta k$ increases with increasing $k_2$, while this increase becomes progressively weaker as either the noise strength $D$ or the phase lag $\beta$ increases. This behavior follows directly from the analytical expressions: the forward critical point $k_{1c}$ is independent of $k_2$, whereas the saddle-node point $k_{1}^{\mathrm{SN}}$ shifts toward smaller values of $k_1$ as $k_2$ increases.

The dark solid symbols represent numerical results for a network of size $N=5000$. Excellent agreement between theory and simulations is observed in the deterministic case ($D=0$; see the orange curves in Fig.~\ref{fig: Fig. 7}a). As the noise strength increases, the discrepancy gradually becomes more pronounced, particularly for larger values of $\beta$ and $k_2$ (see the orange curves in Fig.~\ref{fig: Fig. 7}a). This deviation arises because noise-induced fluctuations can drive the system out of the bistable branch before the deterministic saddle-node bifurcation is reached, leading to an earlier transition than predicted by the deterministic theory.

To further quantify the influence of noise on the stability of the bistable regime, we next analyze the dynamics using Kramers' escape-rate theory.

\textbf{Kramers' escape rate:}Kramers' escape-rate theory ~\cite{RevModPhys.62.251} describes noise-induced transitions between coexisting stable states in a bistable potential. In such a system, two stable fixed points are separated by an unstable barrier. Weak noise perturbs the dynamics within each basin of attraction, whereas sufficiently strong fluctuations enable the system to overcome the potential barrier and transition to the other stable state. The mean escape rate is given by
$\Gamma \propto \exp\!\left(-\frac{\Delta V(k_1,k_2)}{D}\right)$
where $\Delta V$ is the potential barrier height and $D$ is the noise intensity (see Supplementary Sec.~3 for the complete derivation). As the noise strength $D$ increases, the escape rate $\Gamma$ increases exponentially, reducing the mean residence time in each stable state and promoting more frequent transitions between the coexisting stable states ~\cite{RevModPhys.62.251}. In contrast, decreasing $D$ suppresses these transitions and stabilizes the bistable regime. On the other hand, the potential barrier height $\Delta V$ increases with the higher-order interaction strength $k_2$ (see Supplementary Sec.~3). Consequently, larger values of $k_2$ make noise-induced escape less likely, increasing the mean residence time in each stable state and thereby broadening the hysteresis region. This behavior is evident from the comparison of Fig.~\ref{fig: Fig. 3} (for $k_2=10$) and Fig.~\ref{fig: Fig. 4} (for $k_2=20$), where all other parameters and the network size are kept fixed.
Similarly, fig. ~\ref{fig: Fig. 7} shows that, for fixed $k_2$ and $D$, the hysteresis width decreases with increasing phase lag $\beta$. As $\beta$ increases from $\pi/6$ to $\pi/3$, the steady-state order parameter decreases (Eq.~\ref{Eq. 16}), leading to a reduction in the potential barrier height $\Delta V$ (see Supplementary Sec.~3). According to Kramers' escape-rate theory, a lower potential barrier results in a larger escape rate and a shorter mean residence time in the bistable state. Consequently, the system escapes more readily from the bistable branch, leading to a narrower hysteresis region. This behavior is consistent with the numerical results shown in Figs.~\ref{fig: Fig. 3}, \ref{fig: Fig. 4} and \ref{fig: Fig. 6}.

\section{\label{sec:level4}Conclusion}
We investigated the combined effects of phase frustration, Gaussian white noise, and higher-order interactions on synchronization in a globally coupled noisy Kuramoto--Sakaguchi model. Our results show that synchronization is governed by the interplay among these mechanisms. Strong phase frustration favors synchronization when both $1$-simplex and $2$-simplex interactions are repulsive, while sufficiently strong repulsive $2$-simplex coupling can induce coherence even in the presence of weak attractive $1$-simplex interactions. In contrast, under weak frustration, synchronization is promoted when both interactions are attractive, with a sufficiently strong attractive coupling compensating for a weak repulsive coupling of the other interaction type. The system exhibits continuous and discontinuous synchronization transitions, hysteresis, bistability, and noise-induced transitions. Increasing noise suppresses synchronization, smooths the synchronization transition, and progressively shrinks the bistable region, whereas stronger $2$-simplex interactions enlarge the hysteretic region by shifting the backward saddle-node bifurcation while leaving the forward critical point nearly unchanged.
Using the Ott-Antonsen reduction, we derived a low-dimensional amplitude equation that qualitatively predicts the synchronization transitions, critical coupling strengths, bifurcation structure, and bistable region, with good agreement between the reduced theory and direct numerical simulations. The results remain qualitatively unchanged for the representative initial conditions considered and converge toward the analytical predictions with increasing system size, indicating weak finite-size effects. Furthermore, Kramers' escape-rate analysis explains the progressive reduction and eventual disappearance of bistability with increasing noise. Overall, our work provides a unified theoretical framework for understanding synchronization in noisy Kuramoto-Sakaguchi systems with higher-order interactions and offers insights into the collective dynamics of frustrated oscillator networks encountered in physical, biological, and engineered systems.

\section*{Data Availability Statement}
Data sharing is not applicable to this article, as no new data were created or analyzed in this study.

\begin{acknowledgments}
CM acknowledges support from the Anusandhan National Research Foundation (ANRF) India (Grant Numbers SRG/2023/001846 and EEQ/2023/001080) and the INSPIRE-Faculty Grant (No. IFA19-PH248).
\end{acknowledgments}

\section*{Supplementary Material Reference}
Detailed derivations and intermediate steps of the analytical analysis are provided in the supplementary material.

\nocite{*}

\bibliography{apssamp}% Produces the bibliography via BibTeX.

\end{document}